\documentclass[aps,prl,reprint]{revtex4-1}
\usepackage{blindtext}
\usepackage[section]{placeins}

\usepackage{graphicx}
\usepackage{dcolumn}
\usepackage{bm}
\usepackage{amsmath}
\usepackage[mathlines]{lineno}
\modulolinenumbers[5]
\linenumbers\relax 
\begin{document}
\nolinenumbers
\preprint{AAPM/123-QED}

 \title{Ballistic deposition with memory:\\a new universality class of surface growth with a new scaling law}

\author{Ahmed Roman$^{1}$}
\author{Ruomin Zhu$^{4}$}
\author{Ilya Nemenman$^{1,2,3}$} 
\affiliation{$^{1}$Physics Department, Emory University, Atlanta, Georgia, USA\\
$^{2}$ Biology Department, Emory University, Atlanta, Georgia, USA\\
$^{3}$ Initiative for Theory and Modeling of Living Systems, Emory University, Atlanta, Georgia, USA\\
$^{2}$ School of Physics, The University of Sydney, Sydney, NSW, Australia}

\date{\today}

\begin{abstract}
Motivated by recent experimental studies in microbiology, we suggest a modification of the classic ballistic deposition model of surface growth, where memory of a deposition at a site induces more depositions at that site or its neighbors. By studying the statistics of surfaces in this model, we obtain three independent critical exponents: the growth exponent $\beta =5/4$, the roughening exponent $\alpha = 2$, and the new (size) exponent $\gamma = 1/2$. The model requires a modification to the Family-Vicsek scaling, resulting in the dynamical exponent $z = \frac{\alpha+\gamma}{\beta} = 2$. This modified scaling collapses the surface width vs time curves for various lattice sizes. This is a previously unobserved universality class of surface growth that could describe surface properties of a wide range of natural systems.    
\end{abstract}

\maketitle
{\em Introduction.}
Interface growth, and its ensuing roughening, is a paradigmatic nonequilibrium statistical physics process, with applications to many domains of physics \cite{barab}.  Analytical, computational, and experimental studies have shown that the statistics of interface roughness in such processes usually is characterized by one of three well-known universality classes: Poisson, Edwards-Wilkinson (EW), and Kardar-Parisi-Zhang (KPZ) \cite{EW,Kardar}. In the first, interface heights at every point are uncorrelated. In the second, peaks in the interface are smoothed through diffusion. Finally, in the third, nearby sites in the interface help each other grow, resulting in a nonlinear amplification of fluctuations. Competition between the smoothing and the nonlinearity leads to the interface roughness that increases with time and eventually saturates at a system-size dependent value. 

More concretely, we denote the height of a 1-d interface at point $x$ at time $t$ by $h(x,t)$. Then the standard deviation of the interface height defines the interface roughness 
\begin{equation}
w(L,t) = \langle (h(x,t)-\langle h(x,t)\rangle_L)^2\rangle_L^{1/2},
\label{widtheq}
\end{equation} 
and the average here is over a domain of size $L$. Such growth processes are generally characterized by three critical exponents: $\beta$, the {\em growth} exponent, which measures how the roughness grows with time; $\alpha$, the {\em roughness} exponent, which parameterizes the dependence of the roughness of the saturated interface on the system size, and $z$, the dynamical exponent, which relates the time at which the width of the interface stops growing to the system size. The three exponents are related by the celebrated Family-Vicsek dynamical scaling \cite{family}
\begin{align}
w(L,t)&\sim L^{\alpha} f(t/L^z),\quad \mbox{with}\label{eq:FVscaling}\\
f(u) &\propto \bigg\{ \begin{array}{ll}
u^\beta        & u\ll 1 \\
1 & u\gg 1
\end{array} ,  \label{eq:fu}
\end{align}
which results in $z = \alpha/\beta$. 

\begin{figure}[!tbh]
\includegraphics[width=\linewidth]{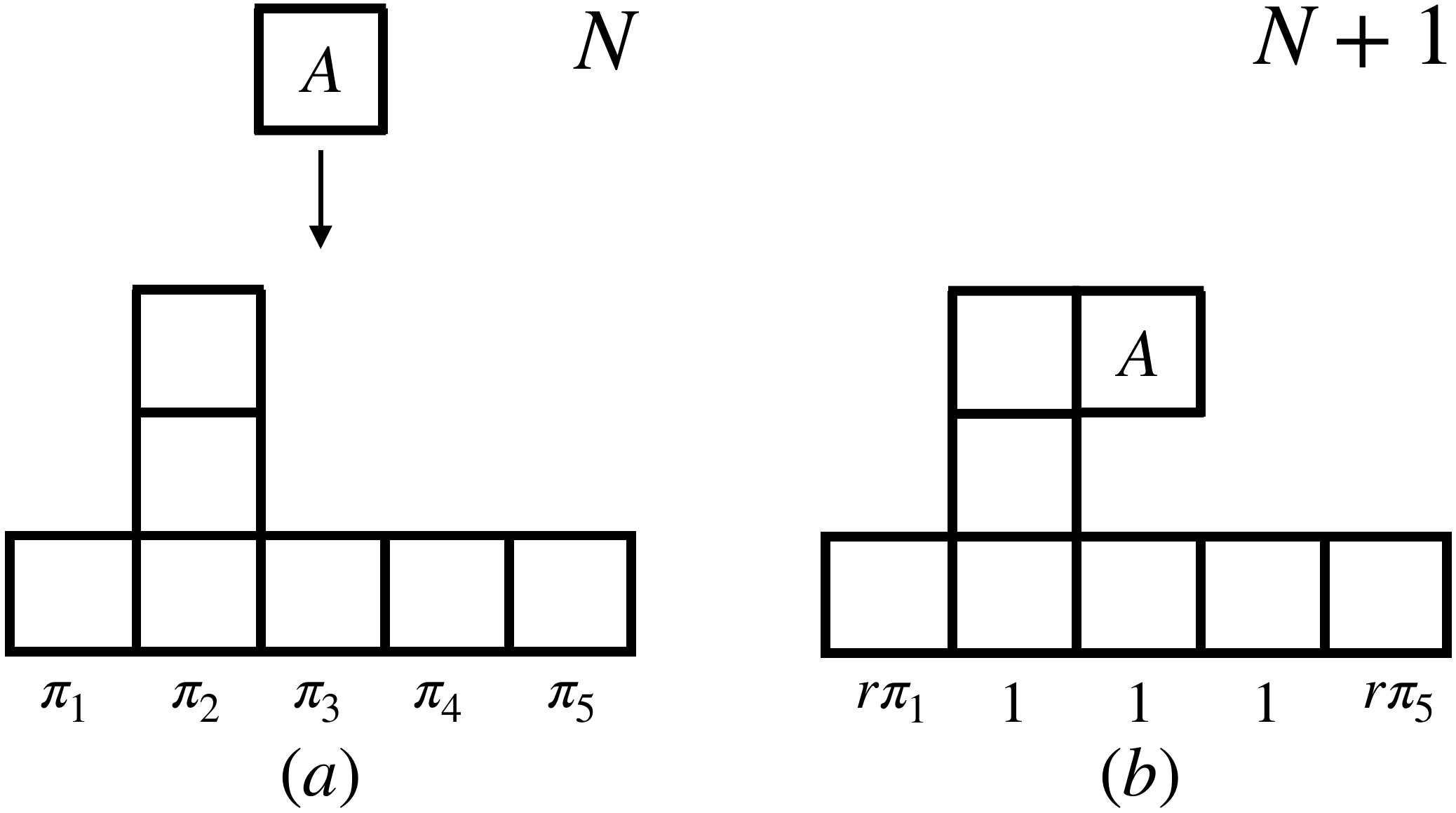}
\caption{\label{fig:epsart}Changes in height and propensity in one time step. The left panel (a) shows the deposition of particle $A$ at step $N$ into site three. The site number increases from left to right. The sites 1-5 have propensities $\pi_1,\cdots, \pi_5$ respectively. Panel (b) indicates the final position of particle $A$ at step $N+1$ as an overhang of column two. The propensities of site three and its two nearest neighbors become one while other site propensities are reduced by factor $r$.}
\label{fig:model}
\end{figure}
\begin{figure*}[!htb]
\includegraphics[width=\textwidth]{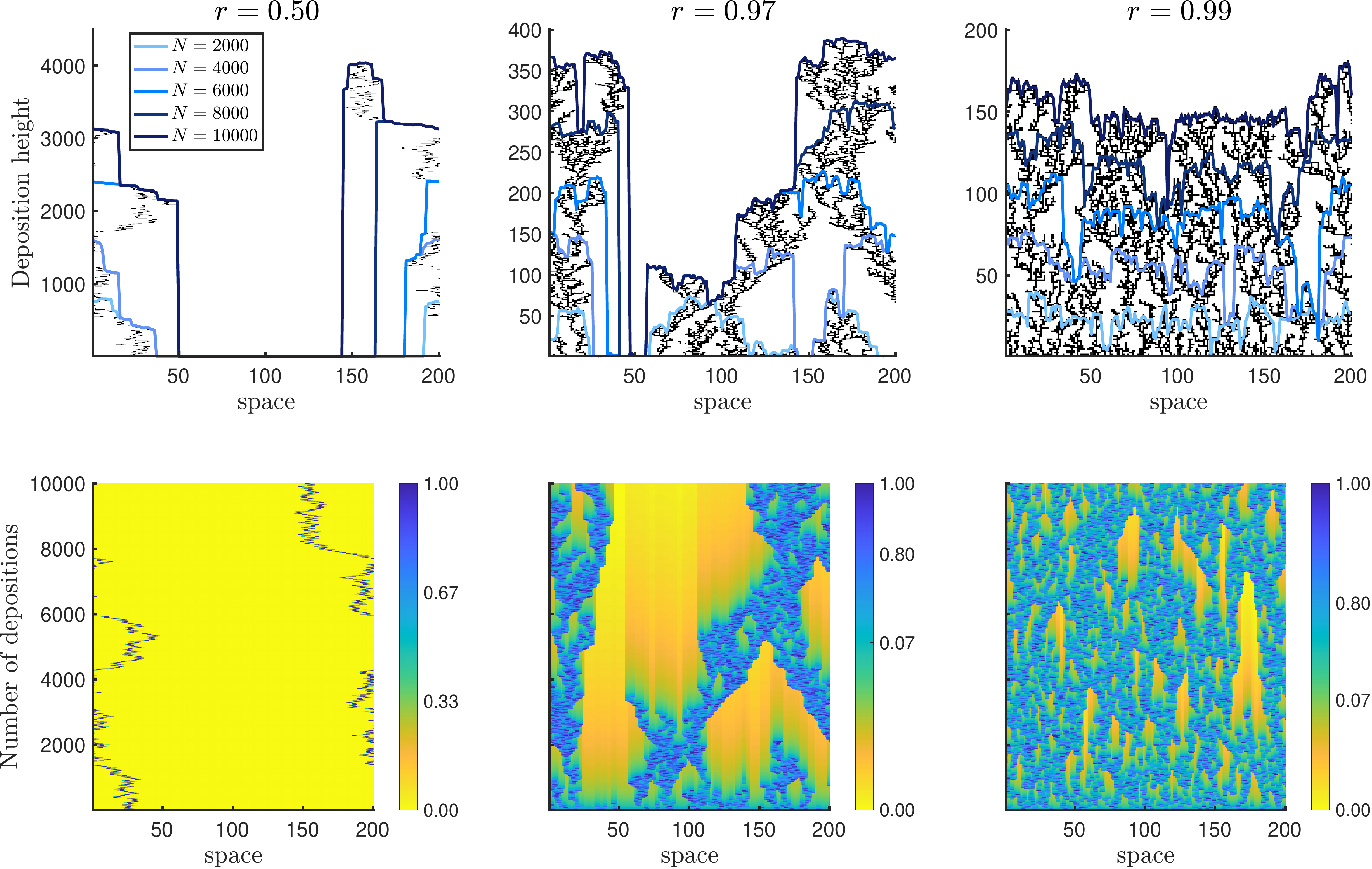}
\caption{\label{fig:propenregimes} Examples of the BDM interfaces. In the top row, the deposition height vs.\ space coordinate is shown for each deposition as a black dot. Different columns are for different  values of $r$. The blue curves show the surface height as a function of the spatial coordinate for different number of depositions, from $N=2000$ (light blue), to $N=10000$ (dark blue). The bottom row shows the propensity, encoded by color, as a function of space and time.}
\end{figure*}
What unites all of these cases is that there is no memory or inertia in the interface growth -- deposition is Markovian in time. This is a reasonable assumption when the interface is built by or from stateless agents. However, when the agents are more complex, such as when they are living cells with a multitude of internal states, such memory-less assumption should be questioned. For example, in cyclic AMP signaling in {\em Dictyostelium discoideum}, which is a classic biological model of collective signaling, collective motility, and development, a spreading wave of cyclic AMP activates a cell, but only if the temporal derivative of the cyclic AMP concentration is positive and large \cite{Wang}. In another example, an action potential propagates in a bacterial film only if a concentration of a previously secreted extracellular potassium has not yet decayed through diffusion \cite{Prindle,Matinez}. All such processes possess memory: the interface at a certain point can grow, but only if it grew here recently. Theory of such interface growth processes with memory is not yet established. In particular, we do not know the relevant critical exponents, how many different universality classes there are, and whether the Family-Viscek scaling is satisfied in such settings. 

Here we develop a model of Ballistic Deposition with Memory (BDM), one of likely many possible extensions of the traditional memoryless surface growth processes, which is inspired by the microbiological systems mentioned above. We derive the critical exponents, and verify them numerically. We show that the process falls into a new universality class with a new scaling law and a new scaling relation. The KPZ universality class is an unstable fixed point in the BDM dynamics. Finally, we discuss the effect of varying memory duration, and show that the standard KPZ interfaces are achieved in a particular limit of the memory parameters. 
 
{\em Model formulation.}
We consider the deposition of particles on a one-dimensional substrate of length $L$. Each site $i$ has a propensity value $0\le \pi_i(N) \le 1$, which determines the probability that the site will receive a particle deposition at step $N$. Initially, all sites are equally likely to receive a deposition; i.e.,  $\pi_i(0)=1$ for all $i$. However, unlike in the ballistic deposition model, if a site $j$ receives a deposition, then the propensity at that site and its nearest neighbors is set to one, while the propensity of all other sites is reduced by a factor $r$, thus reducing the probability of receiving a deposition if no deposition has happened for a long time: \begin{equation}\pi_i(N+1)= \begin{cases} 
      1        & j-1\le i \le j+1 \\
      r \pi_i(N) & \text{otherwise.} 
   \end{cases}
\end{equation} 
Overall, the probability to receive the deposition at site $i$ at the $N$th deposition event is
\begin{equation}
    \mathcal{P} [i,N] = \frac{ \pi_i(N)}{\sum_{i=1}^{L} \pi_i(N)}. \label{prob}
\end{equation} 

\begin{figure*}[!t]
\includegraphics[width=\textwidth]{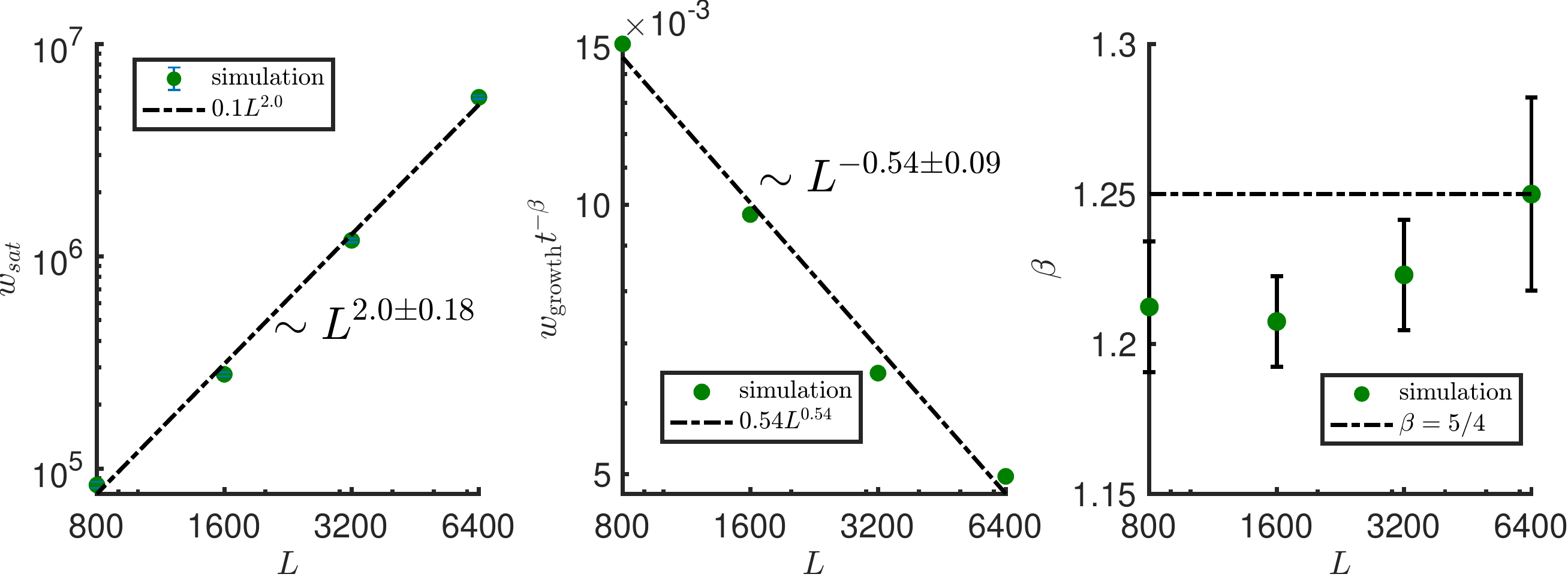}
\caption{\label{fig:gamma}(Left) The saturation width of the BDM interface as a function of the system length $L$. The scaling $L^{\alpha}$, where $\alpha$ is the roughening exponent, obtained from simulation, is shown. (Middle) The scaling of  $w/t^{\beta}$ in the growth regime as function of the system size is $L^\gamma$; $t(N)$ is defined as in     Eq.~(\ref{time}). The fitted scaling with the size exponent $\gamma=0.54\approx 0.5$ is also plotted. (Right) The growth exponent $\beta$ as a function of the system size $L$. The black dash-dotted line at $\beta=5/4$ is the theoretical prediction, in agreement with simulations once finite size effects become negligible. The value $r=0.5$ is used for all subplots.}
\end{figure*}

At step $N$, the height of the interface at site $i$ is $h(i,N)$, with $h(i,0)=0$. After a site $i$ is randomly selected for the deposition according to Eq.~(\ref{prob}), its height increases from $h(i,N)$ to 
\begin{equation}
h(i,N+1) = \max \{h(i-1,N),h(i,N)+1,h(i+1,N)\},
\label{process}
\end{equation}
allowing for overhangs, as in the traditional Ballistic Deposition, cf.~Fig.~\ref{fig:model}. We model the process with periodic boundary conditions, $h(L+1,N) = h(1,N)$. The dynamics of the surface and the propensity are shown in Fig.~\ref{fig:propenregimes} for various values of $r$. For low values of $r$, a single propensity finger moves randomly, causing the deposition sites to follow a random walk, and overhangs form every time the random walk reverts. In the intermediate $r$ regime, multiple propensity fingers move randomly, merge and split. These fingers deposit particles that form shapes, reminiscent of the Diffusion Limited Aggregation \cite{DLA}, though whether the similarity is more than qualitative is unclear. For values of $r\approx1$, many deposition fingers merge into a deposition front, whose fluctuation is KPZ-like (as we will discuss later), but with chasms that have a much lower height, and whose frequency gets lower as $r\to1$.

{\em A random walker.}
We start with the fast propensity decay limit, defined as $r\ll 1/L$ (see {\em Appendix} for details). Here the probability that any site $j$ receives a deposition at step $N+1$ given that 
a non-neighboring site $i$ received a deposition at step $N$ is
\begin{equation} 
\begin{split}
\sum_{\substack{j \not \in \{i-1,i,i+1\} \\ 1\le j \le L}} \mathcal{P} [j,N+1|i,N] & = \sum_{\substack{j \not \in \{i-1,i,i+1\} \\ 1\le j \le L}} \frac{\pi_j(N)}{\sum_{k=1}^L \pi_k(N)}\\
 & < \frac{L r}{3}\ll \frac{1}{3},
 \end{split}
\end{equation}
where we used the bounds $\sum_{k=1}^L \pi_i(N)>\sum_{k \in \{i-1,i,i+1\}} \pi_k(N)=3$ and $\sum_{\substack{j \not \in \{i-1,i,i+1\} \\ 1\le j \le L}} \pi_j(N)<(L-3)r<Lr$ to bound the denominator and numerator, respectively. Therefore, the location $x_N$ of the deposition after $N$ steps is well approximated by a 1-d random walker.

{\em Determining the unit of time.} Up until now, we did not make an explicit distinction between time and $N$, the number of particles deposited, so that $t\sim N$.  This is in contrast to classical models of surface growth, such as ballistic deposition or KPZ, where the time is defined in the units of the mean number of deposited layers, $t\sim N/L$. Adopting the traditional definition would imply that, for small $r$, in one time unit, one deposits $L$ particles on $\sqrt{L}$ sites. The flux per unit time at those $\sqrt{L}$ sites would then be $\sim \sqrt{L}$, which is infinite in the thermodynamic limit, $L\rightarrow \infty$. Here we show that, to avoid this pathology, $t\propto N$ is the only acceptable choice.

We begin by defining the space-averaged propensity $\pi(N) = \frac{1}{L} \sum_{i=1}^{L} \pi_i(N)$,
which can be decomposed into two different contributions. At step $N-1$, $L-3$ sites do not receive a deposition and are not neighbors of the deposition site. The total propensity of those sites at step $N$ is $(L-3)r\pi(N-1)$. The total propensity of the site that receives a deposition and its two neighbors is three. Thus we obtain the recursion relation for $\pi(N)$
\begin{equation}
    \pi(N) = \frac{1}{L}\big [(L-3)r\pi(N-1)+3\big].
    \label{recursion}
\end{equation}
The solution of Eq.~(\ref{recursion}) (see {\em Appendix}) has a characteristic time scale of  
$N_{1/e} = \frac{1}{\ln \frac{L}{(L-3)r}}\approx \frac{1}{\ln \frac{1}{r}}$
depositions,  approaching a steady state space-averaged propensity $\pi_* = 3(L-(L-3)r)^{-1}.$
In the limit $N\gg N_{1/e}$, there are $\mathcal{O}(L\pi_*)$ sites whose probability of receiving a deposition is $\mathcal{O}(1/L\pi_*)$ (see {\em Appendix}), while all other sites have probability zero of receiving a deposition. This implies that the effective lattice length is $L\pi_*$ and motivates a definition of time as
\begin{equation}
    t = \eta \frac{N}{L\pi_*} =\eta  \left(\frac{1}{3}-\frac{L-3}{3L}r\right)N
    \label{time}
    \end{equation}
for any constant $\eta$. In all figures and equations, we define time as in Eq.~(\ref{time}) with $\eta=1$.

{\em Dynamical exponents.}
After $N$ depositions, the deposited particles span $\mathcal{O}(\sqrt{N})$ lattice sites and the random walker has performed  $\mathcal{O}(N)$ reversals, with each reversal increasing the height by 1. Thus the average height of the interface is 
$\langle h(N) \rangle \sim c_1 \cdot 0 \cdot \frac{L-\sqrt{N}}{L}+c_2 \cdot N\cdot \frac{\sqrt{N}}{L}\sim\frac{N^{3/2}}{L} = \lambda_{r,L}^{3/2}t^{3/2}/L $
while the mean squared height is
$\langle h(N)^2 \rangle\sim 
c_3 \cdot0^2 \cdot \frac{L-\sqrt{N}}{L}+c_4\cdot N^2 \cdot \frac{\sqrt{N}}{L}\sim\frac{N^{5/2}}{L}
= \lambda_{r,L}^{5/2}t^{5/2}/L $
for $\lambda_{r,L} =1/(L \pi^*)$ and some constants $c_1,\cdots,c_4$.
The resulting mean width of the interface becomes
\begin{equation}
    w(L,t)\sim \left(\lambda_{r,L}^{5/2}\frac{t^{5/2}}{L}-\delta \lambda_{r,L}^{3/4}\frac{t^{3/4}}{L^2}\right)^{1/2}
    \label{width}
\end{equation}
for some constant $\delta$.

In the regime where $\sqrt{N}\ll L$ (i.e., the random walker has yet to span the system), the width of the interface grows with time (the {\em growth} regime), and its value is dominated by the first term in Eq.~(\ref{width}):
\begin{equation}
w(L,t) \sim \lambda_{r,L}^{5/4} t^{5/4}/\sqrt{L}\propto t^\beta L^{-\gamma}. 
\label{growth}
\end{equation} 
This determines the size exponent $\gamma=1/2$ and growth exponent $\beta=5/4$, which are in excellent agreement with simulation values, cf.~Fig.~\ref{fig:gamma}, despite finite size effects, and even with $r$ outside the regime $r \ll 1/L$. 

In the regime where $\sqrt{N}\gtrsim \mathcal{O}(L)$, the random walker has spanned the lattice, and the surface roughness saturates at  
\begin{equation}
   w_{\rm{sat}}\sim N^{5/4}/\sqrt{L}\sim  (L^2)^{5/4}/\sqrt{L}= L^2,
   \label{saturation}
\end{equation}
This determines the roughness exponent $\alpha= 2$, which agrees with the simulations, Fig.~\ref{fig:gamma}.

\begin{figure}[!t]
\includegraphics[width=\linewidth]{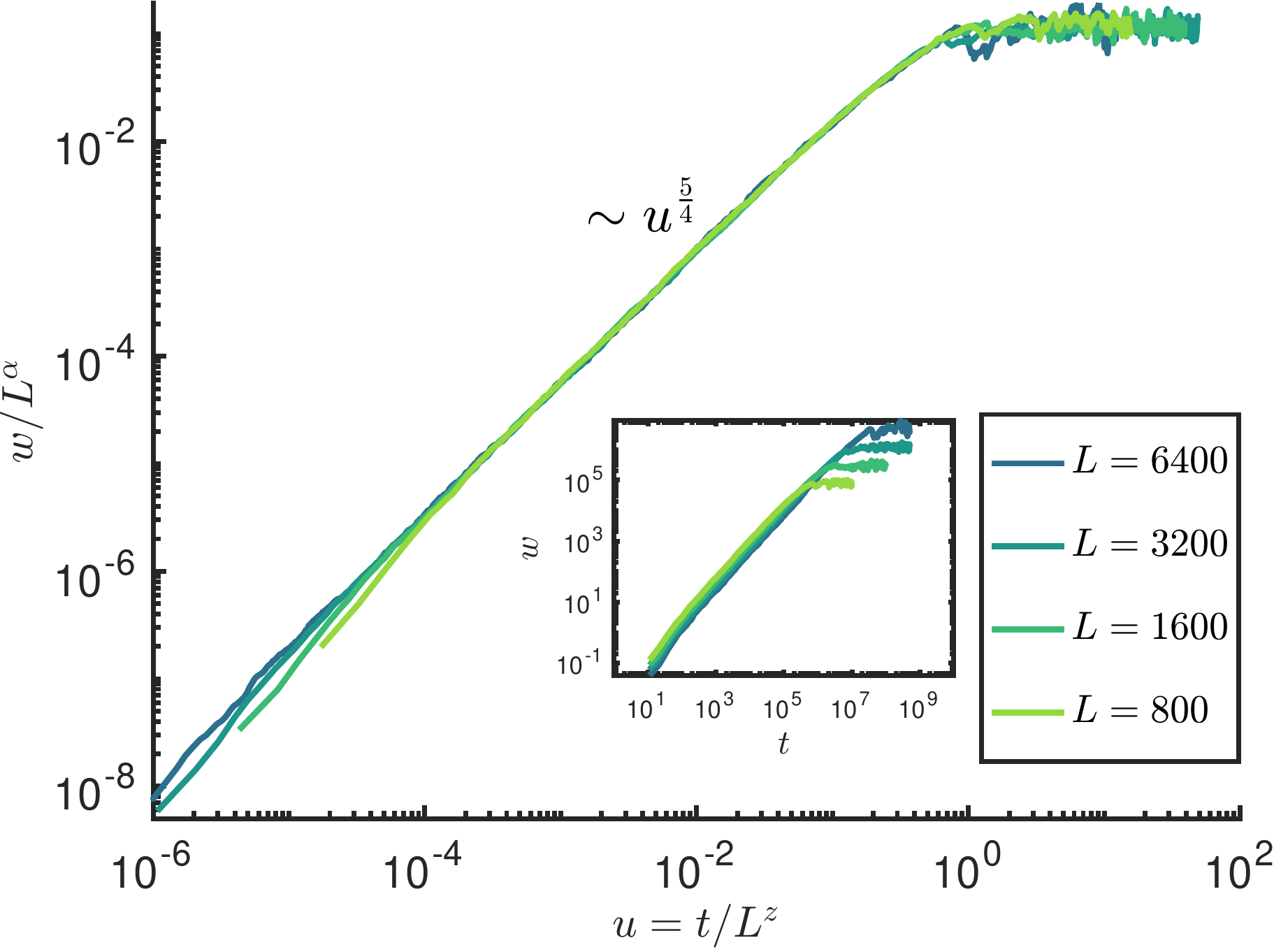}
\caption{\label{fig:width}Interface width $w$ as a function of time for $r=0.5$ and systems of different  lengths, $L$. The time axis is scaled by $L^z = L^2$, and the width axis is scaled by $L^{\alpha}= L^2$, which achieves collapse of all curves. The data come from averaging over $20$ independent runs for all $L$ except $L=6400$, which used $5$ independent runs. The inset shows the bare width as a function of the bare time. }
\end{figure}

{\em Dynamical scaling relation and the scaling law.}
Since the growth and saturation regimes cross at some time $t_{c}$, it follows from Eqs.~(\ref{growth}) and (\ref{saturation}) that $ \lambda_{r,L}^{-\beta} L^{-\gamma} t_{c}^{\beta} \sim L^{\alpha}$. The relation 
$t_{c} \sim \lambda_{r,L}L^{\frac{\alpha+\gamma}{\beta}}= \lambda_{r,L}L^z$ 
determines the scaling law $z = \frac{\alpha+\gamma}{\beta} = 2$ in our model. Note that $\lambda_{r,L}$ has a weak dependence on $L$ such that for $L\gg1$ it is essentially independent of $L$. The scaling relation becomes 
\begin{equation}
    w(L,t)\sim L^{\alpha} f(t/L^{\frac{\alpha+\gamma}{\beta}})
\end{equation}
with $f$ defined as in the Family-Vicsek scaling\cite{family}. Indeed, plotting $w/L^{\alpha}$ against $t/L^{\frac{\alpha+\gamma}{\beta}}$, as in Fig.~\ref{fig:width}, collapses the width vs time curves plotted for various lattice lengths $L$ in the inset of Fig.~\ref{fig:width}. From this, we conclude that there are three independent exponents $\alpha, \beta$ and $\gamma$ that fix the dynamic exponent $z$.

As a newly arriving particle sticks to the surface following Eq.~(\ref{process}), its height is either the same or larger than that of its neighbors. This introduces correlations between neighboring sites. The ensuing height fluctuations spread laterally since particles deposited at nearby sites must have an equal or larger height. This correlation length $\xi_{||}$ can only grow up to the substrate length,  i.e., $\xi_{||}\sim L$ for $t\gg t_c.$ Replacing $L$ by $\xi_{||}$ in $t_c\sim \lambda_{r,L} L^z$, we find that $\xi_{||}\sim \lambda_{r,L}^{-1/z} t_c^{1/z}$ for $t\gg t_c.$ Since $\xi_{||}\sim N^{1/2}$ for  $t\ll t_c$, we see that $\xi_{||}\sim \lambda_{r,L}^{-1/z} t^{1/z}$ holds for $t\ll t_c$ as well. 

\begin{figure}[!t]
\includegraphics[width=\linewidth]{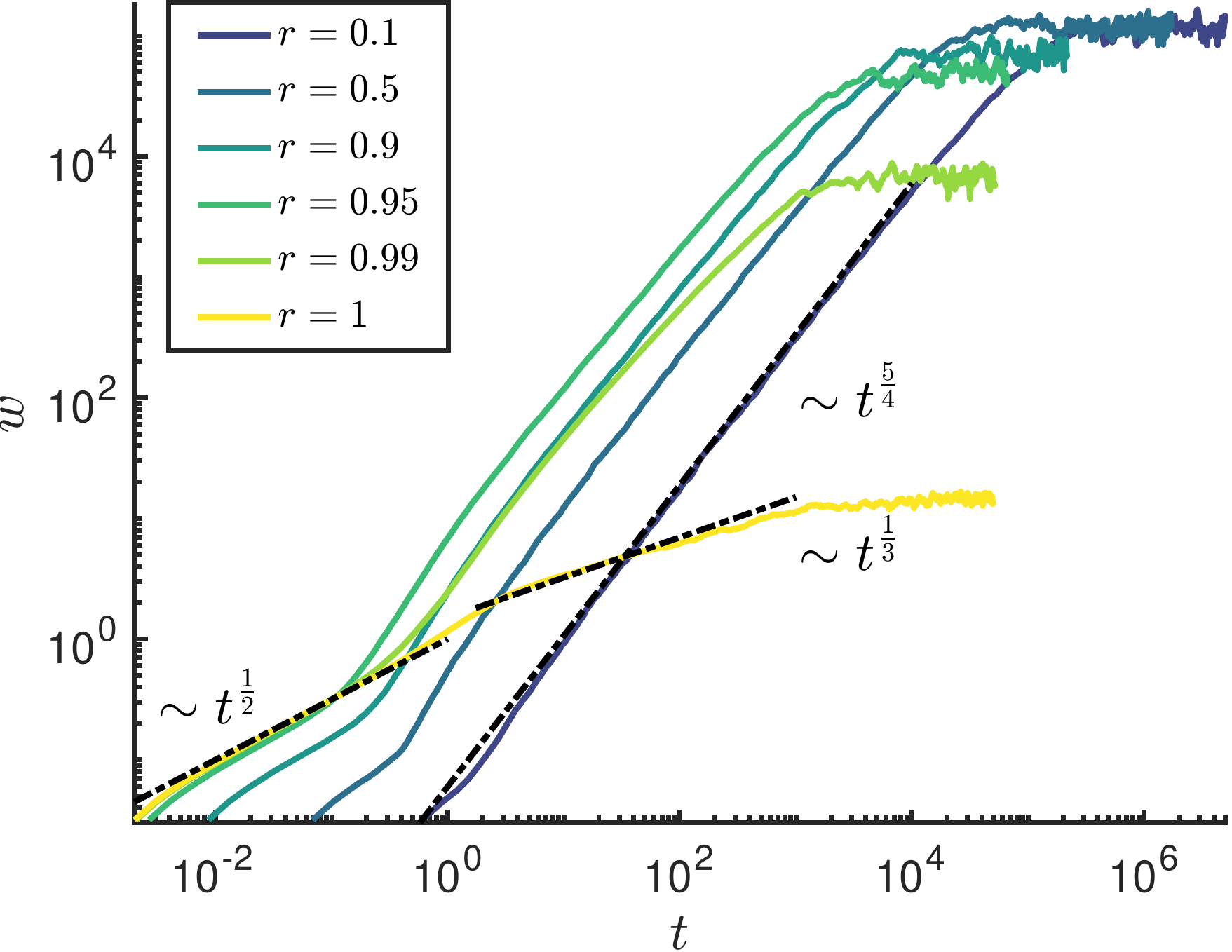}
\caption{Interface width $w$ as a function of time for system size $L=10^3$ and different propensity decay constants $r$. The growth exponent changes from the KPZ value of $\beta = 1/3$ to BDM value $\beta = 5/4$ as the propensity decay rate $r$ deviates from $r = 1$. For $r=1$ and for very small time, the width exhibits the Poisson scaling with $\beta = 1/2$.}
\label{decayconst}
\end{figure}

{\em Varying the memory time scale.}
As $r$ increases, so does the total size of the randomly moving propensity fingers $L\pi_*=\lambda_{r,L}^{-1}$. Increasing $r$ also decreases linearly the time to saturation $t_c$, cf.~Fig.~\ref{decayconst}. In the limit of $r=1$, the KPZ exponents as seen in Fig.~\ref{decayconst} and the standard definition of time $t = N/L$ are recovered. For most of the $r \in[0,1]$ domain, the surface fluctuations are in the new universality class and are not in the KPZ class, cf.~Fig.~\ref{decayconst}. At early times, the Poisson regime dominates the growth with a characteristic scale $t^{1/2}$ followed by the KPZ growth with a scale $t^{1/3}$ (effectively, $r\sim1$) within a moving finger of finite width, and eventual transition to fluctuations with a scale of $t^{5/4}$ (effectively, $r<1$).  For finite $L$, this transition occurs at $r^*(L)<1$. However, in the thermodynamics limit $L\rightarrow \infty$, the transition value $r^*(L)\rightarrow 1$ because the random depositions would cover only a finite part of the lattice. This implies that the KPZ class is an unstable point of the dynamics that occurs only at $r=1$ if $L$ is infinite.    

{\em Conclusion.}
We numerically and analytically studied a model of surface growth with memory. Introduction of memory breaks the temporal locality of the deposition process, so that deposition at the current time is dependent on the history of prior deposition events. This leads to emergence of a new size exponent $\gamma$, which  captures the random walk nature of the deposition process at long times, and to changes in the values of the growth exponent ($\beta = 5/4$) and the roughening exponent ($\alpha=2$). These exponent result in a new scaling law $z= \frac{\alpha+\gamma}{\beta}=2$, which generalized the classical scaling law  $z =\alpha/\beta$. In other words, BDM is a surface growth process that does not belong to the KPZ universality class.

In the standard KPZ and EW universality classes, there is no correlation in the deposition process, so that  the position of the next deposition is independent from the positions of all past depositions. By including the propensity, we introduce such correlations. The ensuing universality class differs from the KPZ and the EW classes, and from their various well-known modifications  \cite{Medina,Sarma,Wolf,Racz,Sun,Lai,Villain}. This is because our novel class changes the dimensionality of the dynamics by introducing additional dynamical variables $\pi$ in addition to $h$.

While our model was inspired by biological systems, it is unclear if the discovered universality class is relevant to them. In order to verify this, it is necessary to explore large spatiotemporal scales that are beyond the typical scales probed in current experiments. For the Dictyostelium discoideum system, tracking the front of the AMP waves on time scales of hours and colonies with radii of order a meter might be necessary to test our predicted exponents. We hope that such experiments will provide exciting new insights in interface growth phenomena. \nocite{*} 

\begin{acknowledgments}
IN thanks Mehran Kardar and Fereydoon Family for useful discussions. This work was supported in part by NSF Grants No.~2010524, 2014173, and by the Simons Foundation.
\end{acknowledgments}

\bibliography{aapmsamp}

\section*{Appendix}

{\em Solving the propensity recursion relation.} Multiplying the recursion relation Eq.~(\ref{recursion}) by the summing factor $(\frac{L-3}{L}r)^{-N-1}$ and summing from $n=0$ to $n=N-1$, we obtain a telescoping sum, which evaluates to
\begin{equation}\pi(N) = \pi(0) \left(\frac{L-3}{L}r\right)^{N-1}+\frac{3}{L} \sum_{n=0}^{N-1}\left(\frac{L-3}{L}r\right)^{n}.
\end{equation}
Rewriting the first term of the above expression and the result of the remaining geometric sum in the exponential form yields   \begin{multline}\pi(N)  = \left(\frac{L}{(L-3)r}-\frac{3}{L-(L-3)r}\right)e^{-N\ln \left(\frac{L}{(L-3)r}\right)} \\+ \frac{3}{L-(L-3)r}.
\end{multline}
From the first term, we obtain the propensity decay time scale 
\begin{equation}N_{1/e} = \frac{1}{\ln\left(\frac{L}{(L-3)r}\right)} \approx \frac{1}{\ln\left(\frac{1}{r}\right)},\; \mbox{for } L\gg 1.
\end{equation}
Thus $\pi(N)$ exponentially decays to 
\begin{equation}
    \pi_*= \frac{3}{L-(L-3)r}
\end{equation}
on the time scale $\mathcal{O}(N_{1/e})$. In the limit $r \rightarrow 1$ and $L\gg 1$,  the time scale $N_{1/e}\rightarrow \infty$. That is,  the propensity remains at the fixed value of $\pi\approx 1$ and no decay occurs, consistent with the regular Ballistic Deposition process. 

{\em Computing deposition probabilities.}
From Eq.~(\ref{prob}), we deduce that if site $j$ receives a deposition at step $N$, then the probability that site $i$ will receive a deposition is 
\begin{align}
\mathcal{P} [i,N+1] &= \frac{ \pi_i(N+1)}{\sum_{i=1}^{L} \pi_i(N+1)}=\frac{\pi_i(N+1)}{L \pi(N+1)}\\
&\approx
\bigg\{ \begin{array}{ll}
\frac{1}{(L-3)r\pi(N)+3},         & \text{for} \; j-1\le i \le j+1 \\
\frac{r \pi(N)}{(L-3)r\pi(N)+3}, & \text{otherwise.}  
\end{array}   
\end{align} 
However, for  $N\gg \frac{1}{\ln(1/r)}$, the propensity $\pi(N)\approx \pi_*=\frac{3}{L-(L-3)r}$
and the probability of a deposition at site $i$ becomes
\begin{equation}
\mathcal{P} [i,N+1]
\approx
\bigg\{\begin{array}{ll} \frac{1}{3}-\frac{L-3}{3L}r,
&\text{for}\; j-1\le i \le j+1 \\
\frac{r}{L}, & \text{otherwise.}  
\end{array}  
\label{probability}
\end{equation} 
As a check, we see that the probability that any site receives a deposition is one
\[\sum_{i=1}^L \mathcal{P} [i,N+1] = (L-3)\frac{r}{L}+3\left(\frac{1}{3}-\frac{L-3}{3L}r\right)=1.\]

Furthermore, using Eq.~(\ref{probability}), we see that
\begin{equation} 
\sum_{\substack{j \not \in \{i-1,i,i+1\} \\ 1\le j \le L}} \mathcal{P} [j,N+1|i,N] = \frac{L-3}{L}r \approx r\ll 1
\end{equation}
for $r\ll 1$. Therefore, the position $x_N$  of the deposition at step $N$ follows an unbiased random walk in this regime.
 
{\em Expanding the random walk regime.}
The steady state propensity $\pi_*$ in practice is not spread out over the entire lattice. Instead $L\pi_* = 3(1-\frac{L-3}{L}r)^{-1}$ sites have propensity almost one, and the rest of the lattice has zero propensity. This motivates the definition of time as $t = \eta \frac{N}{L\pi_*}$ shown in Eq.~(\ref{time}) in the main text. In the limit $L\pi_*\ll L$, the propensity process is made of multiple fingers of cumulative size $L\pi_*$, which all are performing random walks. These fingers dynamically merge and split as particles are deposited randomly. This has the effect that, outside these fingers, the probability of a deposition is zero and hence our dynamical exponents will hold in the significantly larger regime $L\pi_*\ll L$ or, equivalently,
$\frac{3}{1-r}\ll L.$

{\em Extracting exponents from data.}
To estimate the growth exponent $\beta$, we compute the mean width $\langle w(N)\rangle = \frac{1}{n} \sum_{i=1}^{n} w_i(N)$
of $n$ realizations of width $\{w_1(N), \cdots, w_n(N)\}$ obtained from $n$ realizations of the height $\{h_1(N), \cdots, h_n(N)\}$ according to Eq.~(\ref{widtheq}). To ensure that the estimated value of the growth exponent remains in the growth regime and is unaffected by cross-over effects, we limit the range of time used in the estimation to $N \in [3\cdot 10^2,3\cdot 10^4]$. According to Eqs.~\ref{growth} and \ref{saturation}, the mean width in the growth regime is $\langle w(N)\rangle \sim L^{-\gamma} N^{\beta}$. Therefore, linear regression obtains the slopes $\beta$ and $\gamma$ of the plane $\ln w(N)= \beta \ln N - \gamma \ln L + k$  when regressed against $\ln N$ and $\ln L$ respectively. The slopes obtain values of $\beta\approx 1.25 \pm 0.03$ and $\gamma\approx 0.54\pm 0.09$.

To determine the size of the fluctuations around the estimated value of $\beta$ for a fixed substrate length $L$, we use the covariance matrix $\Sigma_{\beta,\Delta}$ of the parameters $\beta$ and $\Delta = -\gamma \ln L+k$. The covariance matrix is $\Sigma_{\beta,\Delta}\approx R^{-1} (R^{-1})^T \frac{|\vec{e_{res}}|^2}{df}$ where $R$ is the triangular factor from a $QR$ decomposition of the Vandermonde matrix of $\ln(N)$, $\vec{e_{res}}$ is the vector of residuals between the data and the fitting line, and $df=2$ is the number of degrees of freedom. The quantity $(\Sigma_{\beta,\Delta})_{\beta,\beta}^{1/2}$ provides the standard deviation on $\beta$. A similar procedure is followed when we regress on $-\gamma \ln L+k$ against $\ln L$ to find $\gamma$ and its standard deviation.

To determine the roughening exponent $\alpha$, we note that $\langle w(t) \rangle \sim L^{\alpha}$ in the saturated regime. To avoid a bias in the estimate of $\alpha$ due to the transition from growth to saturation, we limit the time range used in the estimate to $N \in I_L = [2L^2, 10L^2]$ for lattice length $L$. For a fixed lattice length $L$, the width in the saturation regime fluctuates over the interval $I_L$. In this regime, the mean width is obtained from the relation $\langle w_{\rm{sat}}\rangle_{I_L} =\frac{1}{|I_L|}\sum_{N\in I_L} \langle w(N) \rangle$, where $|I_L|$ is the length of the interval $I_L$ used for the estimate of $w_{sat}$ of the lattice of length $L$. Using linear regression, the value of the slope of the line $\ln\langle w_{\rm{sat}} \rangle = \alpha \ln L + \lambda$  gives $\alpha\approx 2.0\pm 0.18$ as seen in Fig.~\ref{fig:gamma}, in agreement with the value of $\alpha$ obtained analytically.

\end{document}